\documentclass[twocolappendix,appendixfloats,]{emulateapj}

\usepackage[colorlinks = true, linkcolor = blue, urlcolor  = blue, citecolor = blue, anchorcolor = blue]{hyperref}
\usepackage{hyperref}
\usepackage{apjfonts}
\usepackage{rotating}
\usepackage{amsmath}
\usepackage{color}
\usepackage{color}
\usepackage{CJK}

\newcommand{\pivec}{\mbox{\boldmath $\pi$}}
\newcommand{\muvec}{\mbox{\boldmath $\mu$}}
\newcommand{\te}{t_{\rm E}}
\newcommand{\thetae}{\theta_{\rm E}}
\newcommand{\pie}{\pi_{\rm E}}
\newcommand{\pien}{\pi_{{\rm E},N}}
\newcommand{\piee}{\pi_{{\rm E},E}}
\newcommand{\dl}{D_{\rm L}}
\newcommand{\ds}{D_{\rm S}}

\definecolor{darkbrown}{RGB}{139,69,19}

\shorttitle{OGLE-2016-BLG-0156}
\shortauthors{Jung et al.}

\begin{document}

\title{OGLE-2016-BLG-0156: Microlensing Event With Pronounced Microlens-Parallax Effects 
Yielding Precise Lens Mass Measurement}

\author{
Youn~Kil~Jung\altaffilmark{001,101},
Cheongho~Han\altaffilmark{002,104}, Ian~A.~Bond\altaffilmark{003,102}, 
Andrzej~Udalski\altaffilmark{004,100},
Andrew~Gould\altaffilmark{001,005,006,101}, \\
(Leading authors),\\
and \\
Michael~D.~Albrow\altaffilmark{007}, Sun-Ju~Chung\altaffilmark{001,008},  
Kyu-Ha~Hwang\altaffilmark{001}, Chung-Uk~Lee\altaffilmark{001}, Yoon-Hyun~Ryu\altaffilmark{001},
In-Gu~Shin\altaffilmark{009}, Yossi~Shvartzvald\altaffilmark{010}, Jennifer~C.~Yee\altaffilmark{009},
M.~James Jee\altaffilmark{011}, Doeon~Kim\altaffilmark{002},
Sang-Mok~Cha\altaffilmark{001,012}, Dong-Jin~Kim\altaffilmark{001}, Hyoun-Woo~Kim\altaffilmark{001}, 
Seung-Lee~Kim\altaffilmark{001,008}, Dong-Joo~Lee\altaffilmark{001}, Yongseok~Lee\altaffilmark{001,012}, 
Byeong-Gon~Park\altaffilmark{001,008}, Richard~W.~Pogge\altaffilmark{004}\\
(The KMTNet Collaboration),\\
Fumio~Abe\altaffilmark{013}, Richard Barry\altaffilmark{014}, David~P.~Bennett\altaffilmark{014,015},           
Aparna~Bhattacharya\altaffilmark{014,015}, Martin~Donachie\altaffilmark{016}, Akihiko~Fukui\altaffilmark{017},               
Yuki~Hirao\altaffilmark{018}, Yoshitaka~Itow\altaffilmark{013}, Kohei~Kawasaki\altaffilmark{018},              
Iona~Kondo\altaffilmark{018}, Naoki~Koshimoto\altaffilmark{019,020}, Man~Cheung~Alex~Li\altaffilmark{016},               
Yutaka~Matsubara\altaffilmark{013}, Yasushi~Muraki\altaffilmark{013}, Shota~Miyazaki\altaffilmark{018},                
Masayuki~Nagakane\altaffilmark{018}, Cl\'ement~Ranc\altaffilmark{014}, Nicholas~J.~Rattenbury\altaffilmark{016},        
Haruno~Suematsu\altaffilmark{018}, Denis~J.~Sullivan\altaffilmark{021}, Takahiro~Sumi\altaffilmark{018},                
Daisuke~Suzuki\altaffilmark{022}, Paul~J.~Tristram\altaffilmark{023}, Atsunori~Yonehara\altaffilmark{024}\\                    
(The MOA Collaboration),\\
Przemek~Mr\'oz\altaffilmark{004}, Radek~Poleski\altaffilmark{004,005}, Jan~Skowron\altaffilmark{004}, 
Micha{\l}~K.~Szyma\'nski\altaffilmark{004}, Igor~Soszy\'nski\altaffilmark{004}, 
Szymon~Koz{\l}owski\altaffilmark{004}, Pawe{\l}~Pietrukowicz\altaffilmark{004}, 
Krzysztof~Ulaczyk\altaffilmark{004,025}, Micha{\l}~Pawlak\altaffilmark{004} \\
(The OGLE Collaboration) \\   
}

\email{cheongho@astroph.chungbuk.ac.kr}

\altaffiltext{001}{Korea Astronomy and Space Science Institute, Daejon 34055, Republic of Korea} 
\altaffiltext{002}{Department of Physics, Chungbuk National University, Cheongju 28644, Republic of Korea} 
\altaffiltext{003}{Institute of Natural and Mathematical Sciences, Massey University, Auckland 0745, New Zealand}
\altaffiltext{004}{Warsaw University Observatory, Al.~Ujazdowskie 4, 00-478 Warszawa, Poland} 
\altaffiltext{005}{Department of Astronomy, Ohio State University, 140 W. 18th Ave., Columbus, OH 43210, USA} 
\altaffiltext{006}{Max Planck Institute for Astronomy, K\"onigstuhl 17, D-69117 Heidelberg, Germany} 
\altaffiltext{007}{University of Canterbury, Department of Physics and Astronomy, Private Bag 4800, Christchurch 8020, New Zealand} 
\altaffiltext{008}{Korea University of Science and Technology, 217 Gajeong-ro, Yuseong-gu, Daejeon, 34113, Republic of Korea} 
\altaffiltext{009}{Harvard-Smithsonian Center for Astrophysics, 60 Garden St., Cambridge, MA 02138, USA} 
\altaffiltext{010}{IPAC, Mail Code 100-22, Caltech, 1200 E.\ California Blvd., Pasadena, CA 91125, USA}
\altaffiltext{011}{Department of Astronomy, Yonsei University, 50 Yonsei-ro, Seoul 03722, Republic of Korea}
\altaffiltext{012}{School of Space Research, Kyung Hee University, Yongin, Kyeonggi 17104, Korea} 
\altaffiltext{013}{Institute for Space-Earth Environmental Research, Nagoya University, Nagoya 464-8601, Japan}
\altaffiltext{014}{Code 667, NASA Goddard Space Flight Center, Greenbelt, MD 20771, USA}
\altaffiltext{015}{Department of Astronomy, University of Maryland, College Park, MD 20742, USA}
\altaffiltext{016}{Department of Physics, University of Auckland, Private Bag 92019, Auckland, New Zealand}
\altaffiltext{017}{Okayama Astrophysical Observatory, National Astronomical Observatory of Japan, 3037-5 Honjo, Kamogata, Asakuchi, Okayama 719-0232, Japan}
\altaffiltext{018}{Department of Earth and Space Science, Graduate School of Science, Osaka University, Toyonaka, Osaka 560-0043, Japan}
\altaffiltext{019}{Department of Astronomy, Graduate School of Science, The University of Tokyo, 7-3-1 Hongo, Bunkyo-ku, Tokyo 113-0033, Japan}
\altaffiltext{020}{National Astronomical Observatory of Japan, 2-21-1 Osawa, Mitaka, Tokyo 181-8588, Japan}
\altaffiltext{021}{School of Chemical and Physical Sciences, Victoria University, Wellington, New Zealand}
\altaffiltext{022}{Institute of Space and Astronautical Science, Japan Aerospace Exploration Agency, 3-1-1 Yoshinodai, Chuo, Sagamihara, Kanagawa, 252-5210, Japan}
\altaffiltext{023}{University of Canterbury Mt.\ John Observatory, P.O. Box 56, Lake Tekapo 8770, New Zealand}
\altaffiltext{024}{Department of Physics, Faculty of Science, Kyoto Sangyo University, 603-8555 Kyoto, Japan}
\altaffiltext{025}{Department of Physics, University of Warwick, Gibbet Hill Road, Coventry, CV4 7AL, UK} 
\altaffiltext{100}{OGLE Collaboration.}
\altaffiltext{101}{KMTNet Collaboration.}
\altaffiltext{102}{MOA Collaboration.}
\altaffiltext{104}{Corresponding author.}

\begin{abstract}
We analyze the gravitational binary-lensing event OGLE-2016-BLG-0156, for which 
the lensing light curve displays pronounced deviations induced by microlens-parallax 
effects.  The light curve exhibits 3 distinctive widely-separated peaks and we find 
that the multiple-peak feature provides a very tight constraint on the microlens-parallax 
effect, enabling us to precisely measure the microlens parallax $\pie$.  All the peaks 
are densely and continuously covered from high-cadence survey observations using globally 
located telescopes and the analysis of the peaks leads to the precise measurement of the 
angular Einstein radius $\thetae$.  From the combination of the measured $\pie$ and $\thetae$, 
we determine the physical parameters of the lens.  It is found that the lens is a binary 
composed of two M dwarfs  with masses 
$M_1=0.18\pm 0.01\ M_\odot$ and 
$M_2=0.16\pm 0.01\ M_\odot$ located at a distance 
$\dl= 1.35\pm 0.09\ {\rm kpc}$.  
According to the estimated lens mass and distance, 
the flux from the lens comprises an important fraction, $\sim 25\%$, of the blended flux. 
The bright nature of the lens combined with the high relative lens-source motion, 
$\mu=6.94\pm 0.50\ {\rm mas}\ {\rm yr}^{-1}$, 
suggests that the lens can be directly 
observed from future high-resolution follow-up observations. 
\end{abstract}

\keywords{gravitational lensing: micro  -- binaries: general}

\section{Introduction}\label{sec:one}

The microlensing phenomenon occurs by the gravity of lensing objects regardless of their 
luminosity. Due to this property, microlensing provides an important tool to detect 
very faint and even dark objects that cannot be observed by other methods. However, 
it is difficult to conclude the faint/dark nature of the lens just based on the event 
timescale $\te$, which is the only observable related to the lens mass for general 
lensing events,  because the timescale is related to not only the lens mass $M$ but 
also the relative lens-source proper motion $\mu$ and distance to the lens $\dl$ 
and source $\ds$ by
\begin{equation}
\te = {\thetae\over \mu};\qquad
\thetae = \sqrt{\kappa M \pi_{\rm rel}};\qquad
\pi_{\rm rel}={\rm au} \left( {1\over \dl}-{1\over \ds}\right),
\label{eq1}
\end{equation}
where $\kappa=4G/(c^2{\rm au})$ and $\thetae$ is the angular Einstein radius.
In order to reveal the nature of lenses, their masses should be determined.

For the unique determination of the lens mass, it is required to measure two additional 
observables. These observables are the angular Einstein radius $\thetae$ and the 
microlens parallax $\pie$. They are related to the lens mass by \citep{Gould2000}
\begin{equation}
M={\thetae\over \kappa\pie}.
\label{eq2}
\end{equation}
With $\pie$ and $\thetae$, the distance to the lens is also determined by
\begin{equation}
D_{\rm L} = {{\rm au}\over \pie\thetae + \pi_{\rm S}},
\label{eq3}
\end{equation}
where $\pi_{\rm S}={\rm au}/\ds$.

The angular Einstein radius is determined by detecting deviations in lensing light 
curves caused by finite-source effects.  Deviations induced by finite-source effects 
arise due to the differential magnification in which different parts of the source 
surface are magnified by different amount.  For events produced by single mass 
objects, finite-source effects can be detected in the special case in which the 
lens passes over the surface of the source \citep{Witt1994, Gould1994a, Nemiroff1994}.  
However, the ratio of the angular source radius $\theta_*$ to the angular Einstein 
radius $\thetae$ is very small, $\rho~\sim 1/1000$ for events associated with 
main-sequence source stars and $\sim~1/100$ even for events involved with giant 
source stars, and thus the rate of source-crossing events is accordingly very low. 
The chance to detect finite-source effects is relatively much higher for events 
produced by binary objects. This is because a binary lens forms caustics that can 
extend over an important portion of the Einstein ring, and the event with a source 
crossing the caustic exhibits a light curve affected by the finite-source effect 
during the caustic crossing.

In the early-generation lensing surveys that were conducted with a $>1$ day cadence, 
measuring the angular Einstein radius by detecting finite-source effects was 
observationally a challenging task. This is because the duration of the deviation 
induced by finite-source effects is, in most cases, $< 1$ day and thus it was difficult 
to detect the deviation.  The unpredictable nature of caustic crossings also made it 
difficult to cover crossings from high-cadence follow-up observations \citep{Jaroszynski2001}.  
However, with the inauguration of lensing surveys using globally distributed multiple 
telescopes equipped with wide-field cameras, the observational cadence has been dramatically 
increased to $< 1$ hour, making it possible to measure $\thetae$ for a greatly increased 
number of lensing events.

One channel to measure the microlens parallax is simultaneously observing lensing 
events from the ground and in space: `space-based microlens parallax' 
\citep{Refsdal1966, Gould1994b}. The physical size of the Einstein radius for a 
typical lensing event is of the order of au. Then, if space observations are 
conducted using a satellite in a heliocentric orbit, e.g., {\it Deep Impact} 
\citep{Muraki2011} spacecraft or {\it Spitzer Space Telescope} \citep{Dong2007, 
Calchi2015, Udalski2015b}, 
the lensing light curve obtained from the satellite observation will be substantially 
different from that obtained from the ground-based observation.  For events with well 
covered light curves from both the ground and in space, then, the microlens parallax 
can be precisely measured by comparing the two light curves.

Another channel to measure $\pie$ is analyzing deviations induced by 
microlens-parallax effects in lensing light curves obtained from ground-based 
observations: `annual microlens parallax' \citep{Gould1992}. In the single frame 
of Earth, such deviations occur due to the positional change of an observer caused 
by the orbital motion of Earth around the Sun.  For typical lensing events produced 
by low-mass stars, however, the event timescale is several dozen days, which comprises 
a small fraction of the orbital period of Earth, i.e., year, and thus deviations 
induced by the annual microlens-parallax effects are usually very minor.  As a 
result, it is difficult to detect the parallax-induced deviations for general events, 
and even for events with detected deviations, the uncertainties of the measured $\pie$ 
and the resulting lens mass can be considerable.

In this work, we analyze the binary-lensing event OGLE-2016-BLG-0156.  The light 
curve of the event, which is characterized by three distinctive widely-separated 
peaks, exhibits pronounced parallax-induced deviations, from which we precisely 
measure the microlens parallax.  All the peaks are densely covered from continuous 
and high-cadence survey observations and the analysis of the peaks leads to the 
precise measurement of the angular Einstein radius.  We characterize the lens by 
measuring the masses of the lens components from $\pie$ and $\thetae$.

\section{Observation and Data}

The source star of the lensing event OGLE-2016-BLG-0156 is located in the bulge 
field with equatorial coordinates 
$({\rm R.A.},{\rm decl.})_{\rm J2000}=(17:56:36.63, -31:04:40.7)$.  The corresponding 
galactic coordinates are $(l,b)=(359.4^\circ, -3.14^\circ)$. The event was found in 
the very early part of the 2016 bulge season by the Optical Gravitational Lensing 
Experiment \citep[OGLE:][]{Udalski2015a} survey.  The OGLE lensing survey was conducted 
using the 1.3 m telescope of the Las Campanas Observatory, Chile.  The source brightness 
had remained constant, with a baseline magnitude of $I_{\rm base}\sim 18.74$,   until 
the end of the 2015 season since it was monitored by the survey in 2009.  When the 
event was found, the source brightness was already $\sim 0.5$ magnitude brighter than 
the baseline magnitude, indicating that the event started during the $\sim 4$ month 
time gap between the 2015 and 2016 seasons when the bulge field could not be observed 
during the passage of the field behind the Sun.  The OGLE observations were conducted 
at a $\sim 1$--2 day cadence and data were acquired mainly in $I$ band with some $V$ 
band data obtained for the source color measurement.

Two other microlensing survey groups of the Microlensing Observations in Astrophysics 
\citep[MOA:][]{Bond2001, Sumi2003} and the Korea Microlensing Telescope Network 
\citep[KMTNet:][]{Kim2016} independently detected the event. The MOA survey observed 
the event with a $\sim 0.5$ hour cadence in a customized broad $R$ band using the 1.8 m 
telescope of the Mt.~John University Observatory, New Zealand. The event was entitled 
MOA-2016-BLG-069 in the list of MOA transient 
events.\footnote{http://www.massey.ac.nz/~iabond/moa/alert2016/alert.php} 
The KMTNet observations were 
conducted using 3 identical 1.6 m telescopes that are located at the Siding Spring 
Observatory, Australia, Cerro Tololo Interamerican Observatory, Chile, and the South 
African Astronomical Observatory, South Africa. We refer to the individual KMTNet 
telescopes as KMTA, KMTC, and KMTS, respectively. The event, dubbed as KMT-2016-BLG-1709
in the 2016 KMTNet event list\footnote{http://kmtnet.kasi.re.kr/ulens/2016/}, 
was located in the KMTNet BLG01 field toward which observations were conducted with a 
$\sim 0.5$ hour cadence. The field almost overlaps the BLG41 field that was additionally 
covered to fill the gaps between the CCD chips of the BLG01 field. The source happens to 
be located in the gap between the chips of the BLG41 field and thus no data was obtained 
from the field. KMTNet observations were conducted mainly in $I$ band and occasional $V$ 
band observations were carried out to measure the source color.

\begin{figure}
\includegraphics[width=\columnwidth]{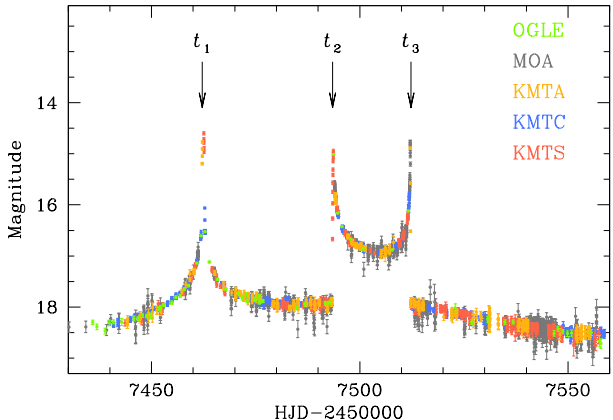}
\caption{
Light curve of OGLE-2016-BLG-0156. The colors of data points correspond to those of 
telescopes, marked in the legend, used for observations. The times marked by arrows 
indicate the centers of the peaks at ${\rm HJD}^\prime={\rm HJD}-2450000\sim 7462.4$ 
($t_1$), 7493.7 ($t_2$), and 7512.1 ($t_3$).
\vskip0.2cm
}
\label{fig:one}
\end{figure}

Photometry of the individual data sets are processed using the codes of the individual 
survey groups. All the photometry codes utilize the difference imaging technique 
developed by \citet{Alard1998}. For the KMTC data set, we additionally conduct 
pyDIA photometry\footnote{The pyDIA code is a python package for performing 
difference imaging and photometry developed by \citet{Albrow2017}. The difference-imaging 
part of this software implements the algorithm of \citet{Bramich2013} with extended delta 
basis functions, enabling independent control of the degrees of spatial variation for 
the differential photometric scaling and differential PSF variations between images.} 
for the source color measurement.  For the use of the multiple data 
sets reduced by different codes, we normalize the error bars of the individual data 
sets using the method described in \citet{Yee2012}.

Figure~\ref{fig:one} shows the light curve of OGLE-2016-BLG-0156. The light 
curve is characterized by three distinct peaks centered at 
${\rm HJD}^\prime={\rm HJD}-2450000 \sim 7462.4$ ($t_1$), 7493.7 ($t_2$), 
and 7512.1 ($t_3$). The peaks are widely separated with time gaps 
$\Delta t_{1-2}\sim 31.3$ days between the first and second peaks and 
$\Delta t_{2-3}\sim 18.4$ days between the second and the third peaks.

\begin{figure}
\includegraphics[width=\columnwidth]{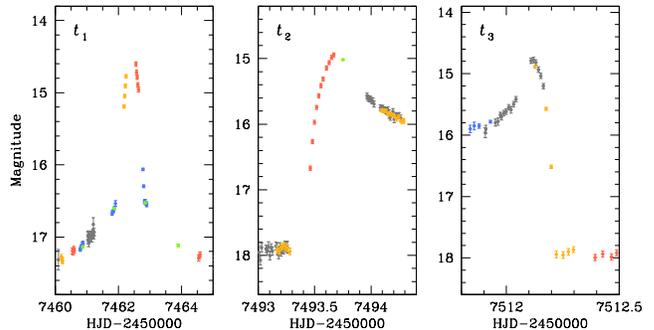}
\caption{
Enlarged view of the three peaks in the lensing light curve.
The locations of the individual peaks in the whole light curve 
are marked by $t_1$, $t_2$, and $t_3$ in Fig.~\ref{fig:one}.
\vskip0.3cm
}
\label{fig:two}
\end{figure}

In Figure~\ref{fig:two}, we present the enlarged views of the individual peaks. 
It is found that the source became brighter by $\gtrsim 3$ magnitudes during very 
short periods of time, indicating that the peaks were produced by the source 
crossings over the caustic. Caustics produced by a binary lens form closed curves 
and thus caustic crossings usually occur in multiples of two.  The region between 
the second and third peaks shows a U-shape pattern, which is the characteristic 
pattern appearing when the source passes inside a caustic, suggesting that the 
pair of the peaks centered at $t_2$ and $t_3$ were produced when the source star 
entered and exited the caustic, respectively. On the other hand, the first peak 
has no counterpart peak. Such a single-peak feature can be produced when the 
source crosses a caustic tip in which the gap between the caustic entrance and 
exit is smaller than the source size.

We note that all the peaks were densely covered. The first peak was covered by the 
combined data sets obtained using the three KMTNet telescopes, the second peak was 
resolved by the OGLE+KMTA data sets, and the last peak was covered by the MOA+KMTS 
data sets. The dense and continuous coverage of all the caustic crossings were 
possible thanks to the coordination of the high-cadence survey experiments using 
globally distributed telescopes.

\begin{figure*}
\epsscale{0.82}
\plotone{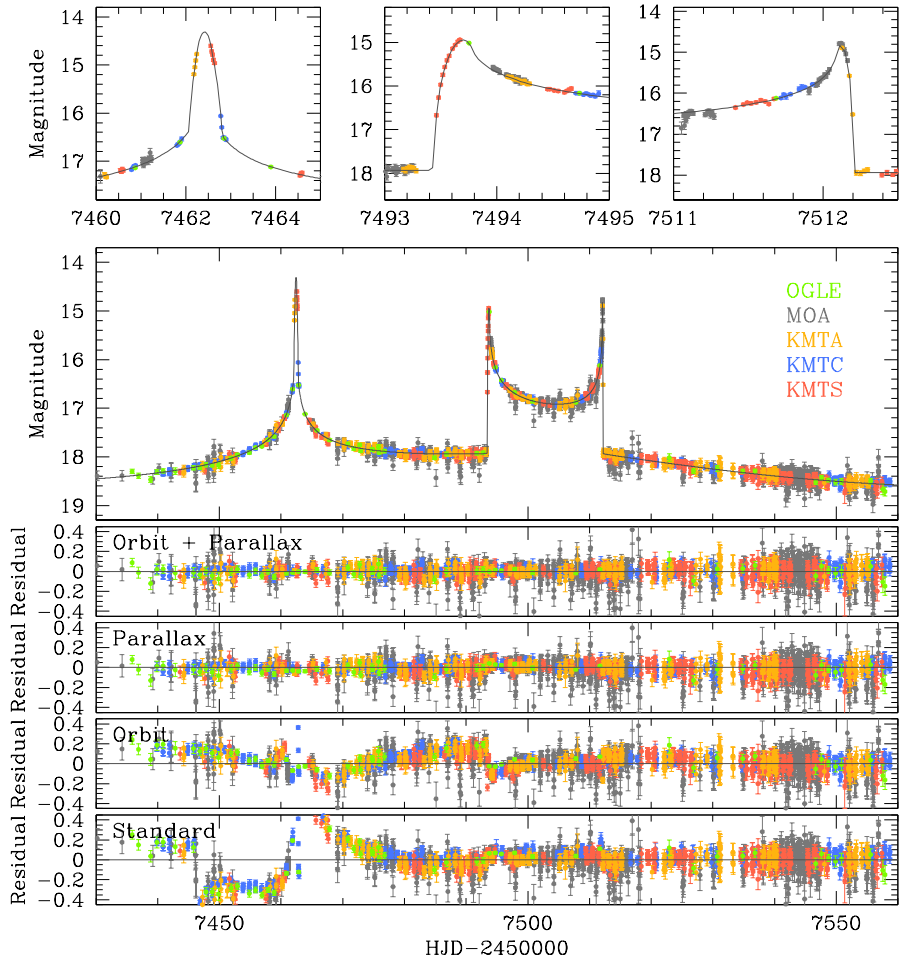}
\caption{
The top two panels show the best-fit model light curve, the curve superposed on the data 
points, obtained by considering both the microlens-parallax and lens-orbital effects. 
The lower four panels compare the residuals of the 4 tested models. The `parallax' 
and `orbit' models are obtained by separately considering the microlens-parallax and 
lens-orbital effects. The `standard' model consider neither of these higher-order 
effects.
\vskip0.3cm
}
\label{fig:three}
\end{figure*}

\section{Modeling Light Curve}

\subsection{Model under Rectilinear Relative Lens-Source Motion}

The caustic-crossing features in the observed light curve indicates that the event is 
likely to be produced by a binary lens and thus we conduct binary-lens modeling of 
the light curve.  We begin by searching for the sets of the lensing parameters that 
best explain the observed light curve under the assumption of the rectilinear 
lens-source motion, wherein the lensing light curve is described by 7 principal parameters. 
The first three parameters $(t_0, u_0, t_{\rm E})$ are identical to those of a 
single-lens events, describing the time of the closest lens-source approach, the 
separation at that time, and the event timescale, respectively. Because a binary lens 
is composed of two masses, one needs a reference position for the lens. We set the 
barycenter of the binary lens as the reference position.  Due to the binary nature 
of the lens, one needs another three parameters $(s, q, \alpha)$, indicating the 
projected binary separation (normalized to $\thetae$), the mass ratio between the 
lens components, and the angle between the binary axis and the source trajectory, 
respectively. The last parameter $\rho$, which represents the ratio of the angular 
source radius $\theta_*$ to the angular Einstein radius, i.e., $\rho=\theta_*/\thetae$ 
(normalized source radius), is needed to describe the deviation of lensing 
magnifications caused by finite-source effects during caustic crossings.

Modeling the light curve is done through a multiple-step process. In the first 
step, we conduct a grid search for the parameters $s$ and $q$ and, for a given 
set of $s$ and $q$, the other parameters are searched for using a downhill approach 
based on the Markov Chain Monte Carlo (MCMC) method. We identify local solutions 
from the $\Delta\chi^2$ maps obtained from this preliminary 
search. In the second step, we refine the individual local solutions first by 
gradually narrowing down the parameter space and then allowing all parameters 
(including the grid parameters $s$ and $q$) to vary. If a satisfactory solution is 
not found from these searches, we repeat the process by changing the initial values 
of the parameters. We refer to the model based on these principal parameters as 
`standard model'.

From these searches, we find that it is difficult to find a lensing model that 
adequately describes the observed light curve.
In the bottom panel of Figure~\ref{fig:three} labeled as `standard', we 
present the residual of the standard model. It shows that the model fit to the 
first peak is very poor although the model relatively well describes the second 
and third peaks.

\subsection{Model with Higher-order Effects}

The difficulty in finding a lensing model that fully explains all the features 
in the observed light curve under the assumption of the rectilinear lens-source 
motion suggests that the motion may not be rectilinear. This possibility is 
further supported by the long duration of the event.

Two major effects cause accelerations in the relative lens-source motion. One is 
the microlens-parallax effect. The other is the orbital motion of the lens: 
lens-orbital effect. We therefore conduct additional modeling considering 
these higher-order effects.

Incorporating the microlens-parallax effect into lensing modeling requires 
two additional parameters of $\pien$ and $\piee$. They represent the two components 
of the microlens-parallax vector $\pivec_{\rm E}$ directed to the north and east, 
respectively. The microlens-parallax vector is related to 
$\pi_{\rm rel}$, $\thetae$, and the relative lens-source proper motion vector 
$\muvec$ by
\begin{equation}
\pivec_{\rm E}={\pi_{\rm rel}\over \thetae}{\muvec \over \mu}.
\label{eq4}
\end{equation}

Considering the lens-orbital effect also requires additional parameters. Under 
the approximation that the positional changes of the lens components induced by 
the lens-orbital effect during the event is small, the effect is described by 
two parameters of $ds/dt$ and $d\alpha/dt$. They represent the change rates of 
the binary separation and the source trajectory angle, respectively.

We conduct a series of additional modeling runs considering the higher-order effects. 
In the `parallax' and `orbit' modeling runs, we separately consider the microlens-parallax 
and lens-orbital effects, respectively. In the `orbit+parallax' modeling run, we 
simultaneously consider both the higher-order effects. For solutions considering 
microlens-parallax effects, it is known that there may exist a pair of degenerate 
solutions with $u_0>0$ and $u_0<0$ due to the mirror symmetry of the source trajectory 
with respect to the binary axis \citep{Smith2003, Skowron2011}.  
We inspect this `ecliptic degeneracy' 
when microlens-parallax effects are considered in modeling.

\begin{figure}
\includegraphics[width=\columnwidth]{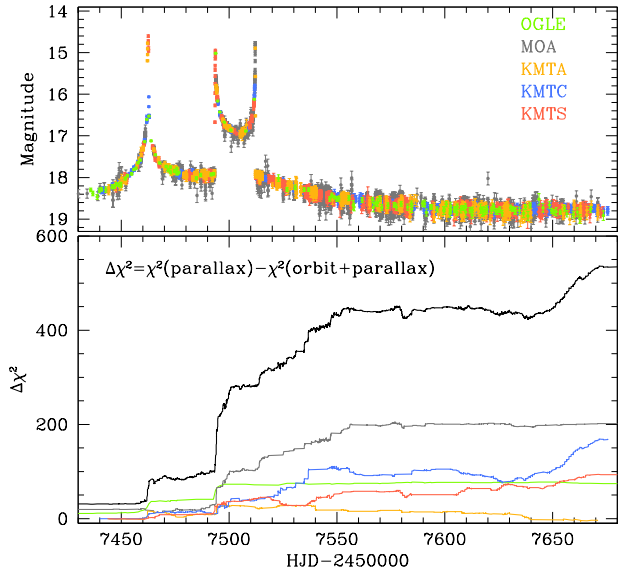}
\caption{
Cumulative distribution of $\chi^2$ difference between the `parallax' and 
`orbit+parallax' models. The light curve in the upper panel is presented to show 
the region of the fit improvement.
}
\label{fig:four}
\end{figure}

\begin{deluxetable}{lcc}
\tablecaption{Comparison of models\label{table:one}}
\tablewidth{240pt}
\tablehead{
\multicolumn{2}{c}{Model}       &
\multicolumn{1}{c}{$\chi^2$}  
}
\startdata                                              
Static            &            &  34436.4   \\
Orbit             &            &  20019.4   \\
Parallax          & $u_0>0$    &  5749.4   \\
--                & $u_0<0$    &  6026.4   \\
Orbit + parallax  & $u_0>0$    &  5215.4   \\
--                & $u_0<0$    &  5221.2     
\enddata                            
\end{deluxetable}

In Table~\ref{table:one}, we list the results of the individual modeling runs in terms 
of $\chi^2$ values of the fits. In order to visualize the goodness of the fits, we also 
present the residuals of the individual models in the lower panels of Figure~\ref{fig:three}. 
For the pair of solutions with $u_0>0$ and $u_0<0$ obtained considering microlens-parallax 
effects, we present the residuals of the solution yielding a better fit.

We compare the fits to judge the importance of the individual higher-order effects. 
From this, it is found that the major features of the light curve, i.e., the three 
peaks, still cannot be adequately explained by the orbital effect alone, although 
the effect improves fit by 
$\Delta\chi^2\sim 14407.0$ 
with respect to the standard 
model. See the residual labeled as `orbit' in Figure~\ref{fig:three}. For the parallax 
model, on the other hand, the fit greatly improves, by 
$\Delta\chi^2\sim 28677.0$, 
and 
all the three peak features are approximately described.  See the residual labeled as 
`parallax' in Figure~\ref{fig:three}. We also find that the fit further improves, by 
$\Delta\chi^2\sim 534.0$ 
with respect to the parallax model, by additionally considering 
the lens-orbital effects. This indicates that although the lens-orbital effect is not the 
prime higher-order effects, it is important to precisely describe the light curve.  Due 
to the relatively minor improvement, it is not easy to see the additional fit improvement 
by the lens-orbital effect from the comparison of the residuals of the `parallax' and 
`orbit+parallax' models.  We, therefore, present the cumulative distribution of $\Delta\chi^2$ 
between the two models as a function of time in Figure~\ref{fig:four}. 
It is found that the fit improves throughout the event and major improvement occurs at the 
first and the second peaks and after the third peak. 
This indicates that the widely separated 
multiple peak features in the lensing light curve help to constrain the subtle higher-order 
effects. 
To check the consistency of the fit improvement, we 
also plot the distributions for the individual data sets.
From the distributions, one finds that the $\chi^2$ improvement shows up in all data sets 
(OGLE, MOA, KMTC, and KMTS) except for the KMTA data set. 
We judge that the marginal orbital signal in the KMTA data set is 
caused by the relatively lower photometry quality than the other 
KMTNet data sets and the resulting smaller number of data points (626 points 
compared to 975 and 1214 points of the KMTS and KMTC data sets, respectively).
We find the ecliptic degeneracy is quite severe although the model with $u_0>0$ is 
preferred over the model with $u_0<0$ by 
$\Delta\chi^2\sim 5.8$.

\begin{figure}
\includegraphics[width=\columnwidth]{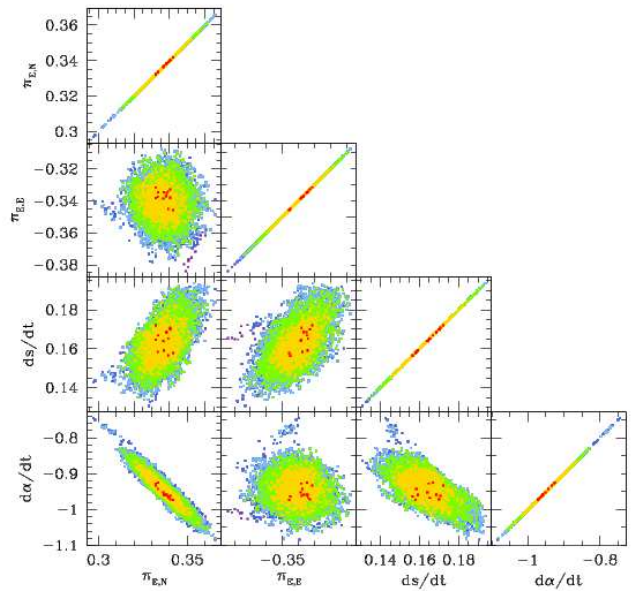}
\caption{
Triangular diagram showing the 
$\Delta\chi^2$ distributions of MCMC points in the planes of the pair of the 
higher-order lensing lensing parameters $\pien$, $\piee$, $ds/dt$, and $d\alpha/dt$.
Points marked in red, yellow, green, cyan, and blue 
represent those with $1\sigma$, $2\sigma$, $3\sigma$, $4\sigma$, and $5\sigma$, respectively.
}
\label{fig:five}
\end{figure}

\begin{deluxetable}{lcc}
\tablecaption{Best-fit lensing parameters\label{table:two}}
\tablewidth{240pt}
\tablehead{
\multicolumn{1}{c}{parameter}       &
\multicolumn{1}{c}{$u_0>0$}      & 
\multicolumn{1}{c}{$u_0<0$}  
}
\startdata                                              
$t_0$ (${\rm HJD}^\prime$) &     7504.730 $\pm$ 0.012       &          7504.644 $\pm$ 0.015     \\
$u_0$                      &     0.083 $\pm$ 0.001          &          -0.085 $\pm$ 0.001       \\
$t_{\rm E}$ (days)         &     68.19 $\pm$ 0.18           &          67.56 $\pm$ 0.39         \\
$s$                        &     0.727 $\pm$ 0.001          &          0.731 $\pm$ 0.002        \\
$q$                        &     0.869 $\pm$ 0.006          &          0.841 $\pm$ 0.006        \\
$\alpha$ (rad)             &     1.408 $\pm$ 0.003          &          -1.408 $\pm$ 0.002       \\
$\rho$ ($10^{-3}$)         &     0.615 $\pm$ 0.013          &          0.620 $\pm$ 0.012        \\
$\pi_{{\rm E},N}$          &     0.334 $\pm$ 0.008          &          -0.347 $\pm$ 0.003       \\
$\pi_{{\rm E},E}$          &     -0.335 $\pm$ 0.011         &          -0.406 $\pm$ 0.012       \\
$ds/dt$ (yr$^{-1}$)        &     0.168 $\pm$ 0.010          &          0.198 $\pm$ 0.013        \\
$d\alpha/dt$ (yr$^{-1}$)   &     -0.958 $\pm$ 0.039         &          1.059 $\pm$ 0.011        \\
$I_{s,{\rm OGLE}}$         &      19.37 $\pm$ 0.01          &          19.37 $\pm$ 0.01         \\   
$I_{b,{\rm OGLE}}$         &      19.77 $\pm$ 0.01          &          19.77 $\pm$ 0.01    
\enddata                            
\tablecomments{${\rm HJD}^\prime={\rm HJD-2450000}$. }
\end{deluxetable}

In Table~\ref{table:two}, we present the lensing parameters of the best-fit models. Because 
the ecliptic degeneracy is severe, we present both the $u_0>0$ and $u_0<0$ 
solutions.  Also presented are the $I$-band magnitudes of the source, $I_{s,{\rm OGLE}}$, 
and the blend, $I_{b,{\rm OGLE}}$, estimated based on the OGLE data.  We note that the 
lensing parameters of the two solutions are roughly in the relation 
$(u_0, \alpha, \pien, d\alpha/dt)\leftrightarrow -(u_0, \alpha, \pien, d\alpha/dt)$
\citep{Skowron2011}. 
Several facts should noted for the obtained lensing parameters. First, the event timescale, 
$t_{\rm E}\sim 68$ days, is substantially longer than typical lensing events with 
$t_{\rm E} \sim 20$ days. Second, the binary parameters $(s,q)\sim (0.73, 0.87)$ indicate 
that the lens is comprised of two similar masses with a projected separation slightly smaller 
than $\thetae$. Third, the normalized source radius $\rho\sim 0.62\times 10^{-3}$ is smaller 
by about a factor $\sim 2.5$ than the value of an event typically occurring on a star with a 
similar stellar type to the source of OGLE-2016-BLG-0156.  
Since $\rho=\theta_*/\thetae$, the 
small $\rho$ value suggests that the angular Einstein radius is likely to be big. Finally, 
the parameters describing the higher-order effects, i.e., $\pien$, $\piee$, $ds/dt$, and 
$d\alpha/dt$, are precisely determined with fractional uncertainties $\sim 2.4\%$, $3.3\%$, 
$6.3\%$, and $4.1\%$, respectively. In Figure~\ref{fig:five}, we present the $\Delta\chi^2$ 
distributions of MCMC points in the planes of the pair of the higher-order lensing parameters.

\begin{figure}
\includegraphics[width=\columnwidth]{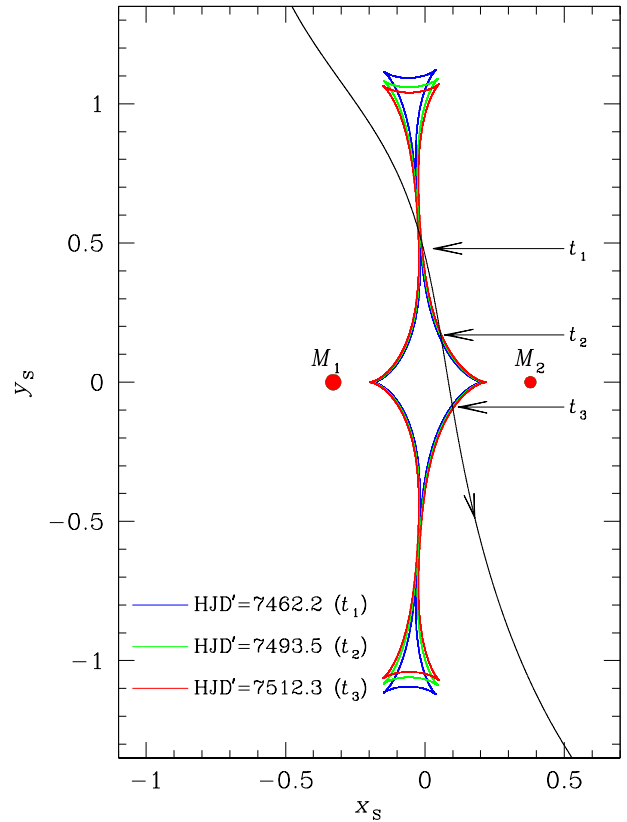}
\caption{
Lens system configuration. The curve with an arrows represents the source trajectory with 
respect to the caustic (closed curve composed of concave segments).  To show the variation 
of the caustic caused by the lens orbital motion, we present caustics at 3 moments 
corresponding to the times of the three peaks in the lensing light curve. We mark the 
positions of source crossings for the individual peaks occurring at $t_1$, $t_2$, and $t_3$. 
The small dots marked by $M_1$ and $M_2$ present the positions of the binary lens components. 
Lengths are scaled to the angular Einstein radius corresponding to the total mass of the lens.
}
\label{fig:six}
\end{figure}

In Figure~\ref{fig:three}, we present the model light curve, which is plotted over the 
data points, of the best-fit solution, i.e., the orbit+parallax model with $u_0>0$. 
In Figure~\ref{fig:six}, we present the corresponding lens-system configuration, showing 
the trajectory of the source with respect to the caustic. When the binary separation $s$ 
is close to unity, caustics form a single closed curve, `resonant caustic', and as the 
separation becomes smaller, the caustic becomes elongated along the direction perpendicular 
to the binary axis and eventually splits into 3 segments, in which one 4-cusp central 
caustic is located around the center of mass and the other two triangular caustics are 
located away from the center of mass.   \citep{Erdl1993, Dominik1999}. For OGLE-2016-BLG-0156, 
the caustic topology corresponds to the boundary between the single closed-curve 
(`resonant') and triple closed-curve (`close') topologies.  The source moved approximately 
in parallel with the elongated caustic, crossing the caustic 3 times at the positions marked 
by $t_1$, $t_2$, and $t_3$. The first peak was produced by the source crossing over the slim 
bridge part of the caustic connecting the 4-cusp central caustic and one of the triangular 
peripheral caustics.  The peak could in principle have been produced by the source star's 
approach to the the right cusp of the upper triangular caustic.  We check this possibility 
and find that it cannot explain the light curve in the region around the first peak.  The 
second and third peaks were produced when the source passed the upper and lower right 
parts of the central caustic, respectively.

\begin{figure}
\includegraphics[width=\columnwidth]{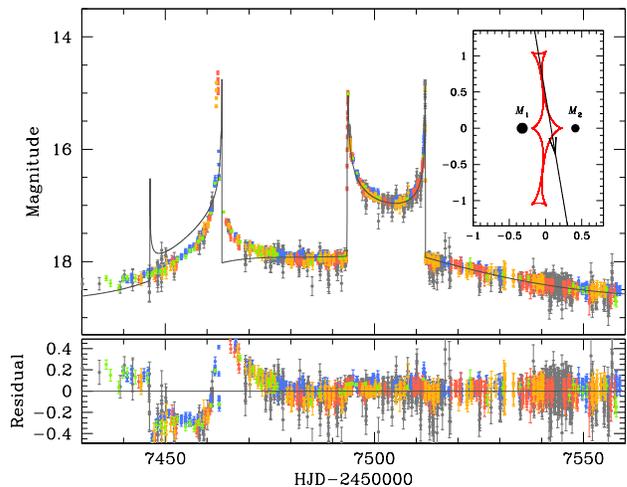}
\caption{
Model light curve (curve superposed on data points) obtained under the assumption 
of the rectilinear relative lens-source motion. The lower panel shows the 
residual from the model. The inset in the upper panel shows the lens system 
configuration corresponding to the model.
\vskip0.2cm
}
\label{fig:seven}
\end{figure}

We note that the well-covered 3-peak feature in the lensing light curve provides a very 
tight constraint on the source trajectory, and thus on the higher-order effects.  To 
demonstrate the high sensitivity of the light curve to the slight change of the source 
trajectory induced by the higher-order effects, in Figure~\ref{fig:seven}, we present 
the model fit of the standard solution and the corresponding lens system configuration. 
One finds that the straight source trajectory without higher-order effects can describe 
the second and third peaks by crossing similar parts of the central caustic to those of 
the solution obtained considering the higher-order effects. However, the extension of 
the trajectory crosses the upper triangular caustic, resulting in a light curve that 
differs greatly from the observed one. The importance of well-covered multiple peaks 
in determining $\pivec_{\rm E}$ was first pointed out by \citet{An2001} and a good 
example was presented by \citet{Udalski2018} for the quintuple-peak lensing event 
OGLE-2014-BLG-0289.

\section{Physical Lens Parameters}

\subsection{Angular Einstein Radius}

For the unique determinations of the mass and distance to the lens, one needs to 
determine $\thetae$ as well as $\pie$. See Equations (\ref{eq2}) and (\ref{eq3}). 
The angular Einstein radius is estimated from the combination of the normalized 
source $\rho$ and the angular source radius $\theta_*$ by $\thetae=\theta_*/\rho$. 
The $\rho$ value is determined from the light curve modeling. Then, one needs to 
estimate $\theta_*$ for the determination of $\thetae$.

\begin{figure}
\includegraphics[width=\columnwidth]{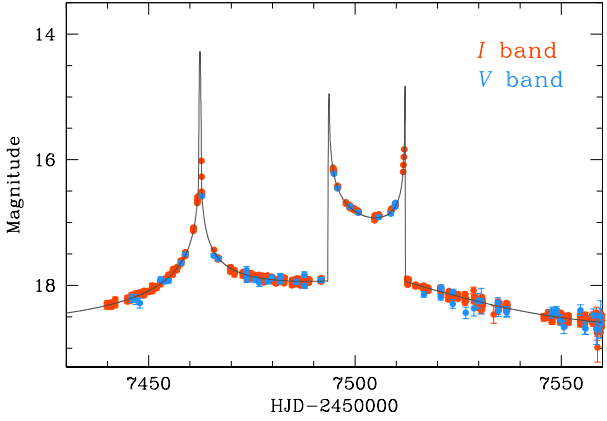}
\caption{
KMTC $I$ and $V$-band data sets processed using the pyDIA photometry code.
The data sets are used for the source color measurement.
}
\label{fig:eight}
\end{figure}

\begin{figure}
\includegraphics[width=\columnwidth]{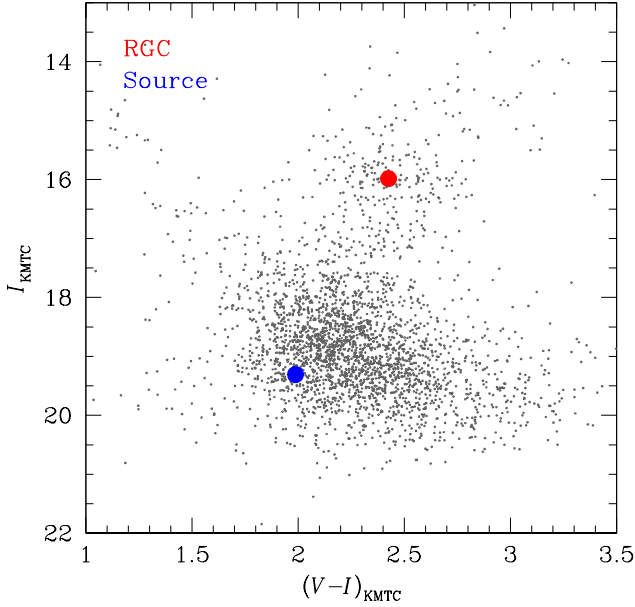}
\caption{
Locations of the source and the centroid of red giant clump (RGC) in the instrumental 
color-magnitude diagram of stars around the source.
The diagram is constructed using the pyDIA photometry of 
KMTC $I$ and $V$-band data.
\vskip0.5cm
}
\label{fig:nine}
\end{figure}

We estimate the angular source radius from the de-reddened color, $(V-I)_0$, 
and brightness, $I_0$.  For this, we first measure the instrumental (uncalibrated) 
source color and brightness from the KMTC $V$ and $I$-band data sets processed using the pyDIA 
photometry. 
We estimate the source color using the regression of the $V$ and $I$-band data sets.
The color can also be estimated by using the model and we find that the source 
color estimated by both ways are consistent.
Figure~\ref{fig:eight} shows the KMTC $I$ and $V$-band data.  In the 
second step, following the method of \citet{Yoo2004}, we calibrate the color and 
brightness of the source using the centroid of the red giant clump (RGC) in the 
color-magnitude diagram as a reference.  In Figure~\ref{fig:nine}, we mark the 
positions of the source, with 
$(V-I, I)=(1.99\pm 0.01, 19.31\pm 0.01)$, 
and the RGC centroid, 
$(V-I, I)_{\rm RGC}=(2.43,15.98)$, in the instrumental color-magnitude diagram.  
With the known de-reddened color and brightness of the RGC centroid, 
$(V-I,I)_{{\rm RGC},0}=(1.06,14.46)$ \citep{Bensby2011, Nataf2013}, combined with 
the measured offsets in color, $\Delta(V-I)=-0.44$, and brightness, $\Delta I=3.33$, 
between the source and the RGC centroid, we estimate that the de-reddened color and 
brightness of the source star are 
$(V-I,I)_{0}=(V-I,I)_{0,{\rm RGC}}+\Delta (V-I,I)=(0.62\pm 0.01, 17.78\pm 0.01)$, 
indicating that 
the source is a turn-off star.  In the last step, we convert the measured $V-I$ 
into $V-K$ using the color-color relation of \citet{Bessell1988} and then estimate 
the angular source radius using the relation between the color and surface 
brightness of \citet{Kervella2004}.  It is estimated that the angular source 
radius is
\begin{equation}
\theta_*=0.79 \pm 0.06\ \mu{\rm as}.
\label{eq5}
\end{equation} 
In addition to the measurement error, the source color estimation is further 
affected by the uncertainty in determining RGC centroid and the differential 
reddening of the field.
\citet{Bensby2013} showed that for lensing events in the fields with 
well defined RGCs, the typical error in the source color estimation is 
about 0.07 mag. 
We, therefore, estimate the errorbar of $\theta_*$ by considering this 
additional error.

The estimated angular Einstein radius is 
\begin{equation}
\thetae= 1.30 \pm 0.09\ {\rm mas}.
\label{eq6}
\end{equation}
For a typical lensing event produced by a low-mass star ($\sim 0.3~M_\odot$) 
located halfway between the source and observer ($D_{\rm L}\sim 4~{\rm kpc}$), 
the angular Einstein radius is 
$\thetae=\sqrt{\kappa M \pi_{\rm rel}}
\sim 0.55~{\rm mas}~(M/0.3~M_\odot)^{1/2}$.
Then the estimated angular Einstein radius is $\gtrsim 2$ times 
bigger than the 
value of a typical lensing event.  This is expected from the small value of the 
normalized source radius. Combined with the event timescale, the relative 
lens-source proper motion in the geocentric frame is estimated by
\begin{equation}
\mu_{\rm geo} = {\thetae\over\te} =
6.94 \pm 0.69\ {\rm mas}\ {\rm yr}^{-1}.
\label{eq7}
\end{equation}
The corresponding proper motion in the heliocentric frame is estimated by
\begin{equation}
\mu_{\rm helio} =
\left\vert
\mu_{\rm geo} {\pivec_{\rm E}\over \pie} +
{\bf v}_{\oplus,\perp}{\pi_{\rm rel} \over {\rm au}}\right\vert =
5.94 \pm 0.43\ {\rm mas}\ {\rm yr}^{-1}.
\label{eq8}
\end{equation}
Here ${\bf v}_{\oplus,\perp}=(v_{\oplus,\perp,N}, v_{\oplus,\perp,E})=
(3.1,17.5)\ {\rm km}\ {\rm s}^{-1}$ denotes the 
projected velocity of Earth at $t_0$.

In Table~\ref{table:three}, we summarize the estimated values of the angular 
Einstein radius, relative lens-source proper motion (in both geocentric and 
heliocentric frames), and the direction of the relative motion, i.e., 
$\phi=\tan^{-1} (\mu_{{\rm helio},E}/\mu_{{\rm helio},N})$.  We also present 
the quantities resulting from the $u_0<0$ solution.  The obtained quantities are 
slightly different from those of the $u_0>0$ solution due to the slight differences 
in $\rho$, $\pien$, and $\piee$.

\begin{deluxetable}{lcc}
\tablecaption{Einstein radius and Proper Motion\label{table:three}}
\tablewidth{240pt}
\tablehead{
\multicolumn{1}{c}{Quantity}       &
\multicolumn{1}{c}{$u_0>0$}        &
\multicolumn{1}{c}{$u_0<0$}         
}
\startdata                                              
$\thetae$                            &   $1.30 \pm 0.09$  &   $1.28 \pm 0.09$    \\
$\mu_{\rm geo}$ (mas yr$^{-1}$)      &   $6.94 \pm 0.50$  &   $6.91 \pm 0.50$    \\
$\mu_{\rm helio}$ (mas yr$^{-1}$)    &   $5.94 \pm 0.43$  &   $4.88 \pm 0.35$    \\
$\phi$                               &   $334^\circ$      &   $215^\circ$        
\enddata                            
\end{deluxetable}
\smallskip

\subsection{Mass and Distance}

With the measured $\pie$ and $\thetae$, the masses of the individual lens components are 
determined as
\begin{equation}
M_1=
0.18 \pm 0.01\ M_\odot,
\label{eq9}
\end{equation}
and
\begin{equation}
M_2=qM_1=
0.16 \pm 0.01\ M_\odot.
\label{eq10}
\end{equation}
It is estimated that the lens is located at a distance
\begin{equation}
\dl = 
1.35\pm 0.09\ {\rm kpc}.
\label{eq11}
\end{equation}
The determined masses and distance indicates that the lens is a binary 
composed of two M dwarfs located in the disk.  The projected separation 
between the lens components is
\begin{equation}
a_\perp = s\thetae\dl =  
1.28 \pm 0.09\ {\rm au}.
\label{eq12}
\end{equation}
We note that the $u_0<0$ solution yields similar lens parameters.
In Table~\ref{table:four}, we list the physical lens parameters for both 
the $u_0>0$ and $u_0<0$ solutions.

We check the validity of the solution by estimating the projected kinetic-to-potential 
energy ratio. We compute the ratio from the physical lens parameters of $M=M_1+M_2$ 
and $a_\perp$ and the measured lensing parameters of $s$, $\alpha$, $ds/dt$, 
and $d\alpha/dt$ by
\begin{equation}
\left( {{\rm KE}\over {\rm PE}}\right)_\perp =
{ (a_\perp/{\rm au})^3\over 8\pi^2(M/M_\odot) }
\left[
\left( {1\over s}{ds/dt\over {\rm yr}^{-1}} \right)^2 +
\left( {d\alpha/dt \over {\rm yr}^{-1}}\right)^2 
\right].
\label{eq13}
\end{equation}
In order for the lens system to be a gravitationally bound system, 
the solution should satisfy the condition of
$({\rm KE}/{\rm PE})_\perp \leq {\rm KE}/{\rm PE} \leq 1.0$,
where ${\rm KE}/{\rm PE}$ denotes the intrinsic energy ratio.
The estimated ratio $({\rm KE}/{\rm PE})_\perp \sim 0.08$ satisfies
this condition. The low value of the ratio suggests that the binary
components are aligned along the line of sight.

\subsection{Lens Brightness}

Although the lens components are M dwarfs, they are located at a close distance, 
and the flux from the lens can comprise a significant portion of the blended flux, 
e.g., OGLE-2017-BLG-0039 \citep{Han2018}.  To check this possibility, we estimate 
the expected brightness of the lens.  The stellar types of the lens components are 
about M4.5V and M5.0V with absolute $I$-band magnitudes of $M_{I,1}\sim 10.5$ and 
$M_{I,2}\sim 11.0$ for the primary and companion, respectively, resulting in the 
combined magnitude $M_I\sim 10.0$.  With the known distance to the lens, the 
de-reddened $I$-band magnitude is then $I_{L,0}=M_I + 5\log\dl -5 \sim 20.6$.  
From the OGLE extinction map \citep{Nataf2013}, the total $I$-band extinction 
toward the source is $A_{I,{\rm tot}}\sim 1.48$.  Assuming that about half of 
the total extinction is caused by the dust and gas located in front of the lens, 
i.e., $A_I\sim 0.7$, the expected brightness of the lens is
\begin{equation}
I_{\rm L} = I_{{\rm L},0}+A_I \sim 
21.3.
\label{eq14}
\end{equation}  
Compared to the brightness of the blend, $I_{b,{\rm OGLE}}\sim 19.8$,
it is found that the the flux from the lens comprises an important fraction,
$\sim 25\%$, of the blended light. 
We note that the color constraint of the blended light cannot be used because the 
uncertainty of the $V$-band blend flux measurement is bigger than the flux itself.

\begin{deluxetable}{lcc}
\tablecaption{Physical lens parameters\label{table:four}}
\tablewidth{240pt}
\tablehead{
\multicolumn{1}{c}{Parameter}      &
\multicolumn{1}{c}{$u_0>0$}        &
\multicolumn{1}{c}{$u_0<0$}  
}
\startdata                                              
$M_1$ ($M_\odot$)             &   $0.18 \pm 0.01$  &  $0.16 \pm 0.01$    \\ 
$M_2$ ($M_\odot$)             &   $0.16 \pm 0.01$  &  $0.13 \pm 0.01$    \\  
$\dl$ (kpc)                   &   $1.35 \pm 0.09$  &  $1.24 \pm 0.08$    \\ 
$a_\perp$    (au)             &   $1.28 \pm 0.09$  &  $1.16 \pm 0.08$    \\   
$({\rm KE}/{\rm PE})_\perp$   &   0.08             &  0.08
\enddata                            
\end{deluxetable}
\smallskip

The bright nature of the lens combined with the high relative lens-source proper 
motion suggests that the lens can be directly observed from high-resolution 
follow-up observations.  
For the case of the lensing event OGLE-2005-BLG-169, 
the lens was resolved from the source on the Keck AO images when they were separated 
by $\sim 50$ mas after $\sim 8$ years after the event \citep{Batista2015}.  By applying 
the same criterion, the lens and source of OGLE-2016-BLG-0156 can be resolved if 
similar follow-up observations are conducted $\sim 6$ years after the event, i.e., 
after 2022. 
For the case of the another lensing event OGLE-2012-BLG-0950, 
\citet{Bhattacharya2018} resolved the source and lens using Keck and the 
{\it Hubble Space Telescope} when they were separated by $\sim 34$ mas. 
According to this criterion, then, 
the source and lens of this event would be resolved in 2022.

Because follow-up observations are likely to to be conducted in near infrared bands, we estimate 
the expected $H$-band brightness of the lens.  The absolute $H$-band magnitudes of 
the individual lens components are 
$M_H\sim 8.2$ and 8.7 resulting in the combined brightness of $M_{H,0}\sim 7.7$.  
With $A_I=A_{I,{\rm tot}}/2\sim 0.7$ and 
$E(V-I)=E_{\rm tot}(V-I)/2\sim 0.6$ from \citet{Nataf2013} and adopting the relation 
$A_H\sim 0.108 A_V\sim 0.14$ of \citet{Nishiyama2008}, we estimate that $H$-band 
brightness of the lens is 
\begin{equation}
H_{\rm L} = M_{H,0} + A_H + 5\log \dl -5 
\sim 17.9.
\label{eq15}
\end{equation}
The $H$-band brightness of the source is 
\begin{equation}
H_{\rm S}=I_{\rm S}-E(I-H)-(I-H)_0 \sim 17.5,
\label{eq16}
\end{equation}
which is similar to that of the the lens.
When the lens brightness is similar to the brightness of the source,
the lens and source can be better resolved as demonstrated for events 
OGLE-2005-BLG-169 \citep{Bennett2015}, 
MOA-2008-BLG-310 \citep{Bhattacharya2017}, and
OGLE-2012-BLG-0950 \citep{Bhattacharya2018}.

\section{Conclusion}

We analyzed a binary microlensing event OGLE-2016-BLG-0156.  We found that 
the light curve of the event exhibited pronounced deviations induced by 
higher-order effects, especially the microlens effect. It is found that the 
multiple-peak feature provided a very tight constraint on the microlens-parallax 
measurement.  In addition, the good coverage of all the peaks from the combined 
survey observations allowed us to precisely measure the angular Einstein radius.  
We uniquely determined the physical lens parameters from the measured values of 
$\pie$ and $\thetae$ and found that the lens was a binary composed of two M dwarfs 
located in the disk.  
We also found that the flux from the lens comprises an important fraction 
of the blended flux.
The bright nature 
of the lens combined with the high relative lens-source motion suggested that the lens 
could be directly observed from high-resolution follow-up observations.

\acknowledgments
Work by CH was supported by the grant (2017R1A4A1015178) of National Research Foundation of Korea.
Work by AG was supported by US NSF grant AST-1516842.
Work by IGS and AG were supported by JPL grant 1500811.
AG received support from the
European Research Council under the European Union's
Seventh Framework Programme (FP 7) ERC Grant Agreement n.~[321035].
The MOA project is supported by JSPS KAKENHI Grant Number JSPS24253004,
JSPS26247023, JSPS23340064, JSPS15H00781, and JP16H06287.
YM acknowledges the support by the grant JP14002006.
DPB, AB, and CR were supported by NASA through grant NASA-80NSSC18K0274. 
The work by CR was supported by an appointment to the NASA Postdoctoral Program at the Goddard 
Space Flight Center, administered by USRA through a contract with NASA. NJR is a Royal Society 
of New Zealand Rutherford Discovery Fellow.
The OGLE project has received funding from the National Science Centre, Poland, grant
MAESTRO 2014/14/A/ST9/00121 to AU.
This research has made use of the KMTNet system operated by the Korea
Astronomy and Space Science Institute (KASI) and the data were obtained at
three host sites of CTIO in Chile, SAAO in South Africa, and SSO in
Australia.
We acknowledge the high-speed internet service (KREONET)
provided by Korea Institute of Science and Technology Information (KISTI).

\end{document}